\documentclass[12pt,citesort]{article}
\usepackage{amsmath,amssymb,graphicx,url}
\newcommand{\ignore}[1]{}
\newcommand\be{\begin{equation}}
\newcommand\ee{\end{equation}}
\newcommand\bea{\begin{eqnarray}}
\newcommand\eea{\end{eqnarray}}

\begin{document}
\title{
Saturation and Confinement: Analyticity, Unitarity and AdS/CFT Correspondence
}
\author{Richard  C. Brower\footnote{Physics Department,
Boston University, Boston MA 02215},
Marko Djuri\'c\footnote{Physics Department, Brown University,
Providence, RI 02912},  and  Chung-I Tan\footnote{Physics Department, Brown University,
Providence, RI 02912. Talk presented at ISMD 2008, 15-20 Sept. 2008.}
}
\maketitle
\begin{abstract}
In  $1/N_c$ expansion, analyticity and crossing lead to crossing even and odd ($C=\pm 1$) vacuum exchanges  at high-energy,  the {\em Pomeron} and  the {\em Odderon}. We discuss how, using {\em String/Gauge duality}, these can be identified  with a  reggeized {\em  Graviton} and  the anti-symmetric {\em Kalb-Ramond fields} in $AdS$ background. With confinement, these Regge singularities interpolate with glueball states. We also discuss unitarization   based on eikonal sum in $AdS$.
\end{abstract}

\section{Forward Scattering, Gauge/String Duality and Confinement}

The subject of near-forward high energy scattering for hadrons has a long history,
predating both QCD and string theory.  We focus here on the recent developments in this
subject based on Maldacena's weak/strong duality, relating Yang-Mills
theories to string theories in (deformed) Anti-de Sitter space~\cite{Brower:2006ea,Brower:2007qh,Brower:2007xg,BDT,Cornalba:2006xk,Cornalba:2006xm,Cor}.  For conformally invariant gauge theories, the metric of the dual
string theory is a product, $AdS_5 \times W$,  $
ds^2 =\left( \frac{r^2} {R^2}\right) \eta_{\mu\nu} {dx^\mu dx^\nu} +\left(\frac{R^2} {r^2} \right)  {dr^2} + ds^2_W\ ,$
where $0 <r < \infty$.  For the dual to ${\cal N}=4$
supersymmetric Yang-Mills theory  the AdS radius
$R$ is
$R^2 \equiv\sqrt{\lambda} \alpha'= (g_{\rm YM}^2 N)^{1/2} \alpha' \ ,$
and $W$ is a 5-sphere of this same radius.  We will ignore fluctuations over $W$ and also  assume  that $\lambda \gg 1$, so that the spacetime
curvature is small on the string scale, and $g^2_{YM} \ll 1$ so
that we can use string perturbation theory.  

\begin{figure}[h]
\quad
\includegraphics[height=0.15 \textwidth,width=0.35\textwidth]{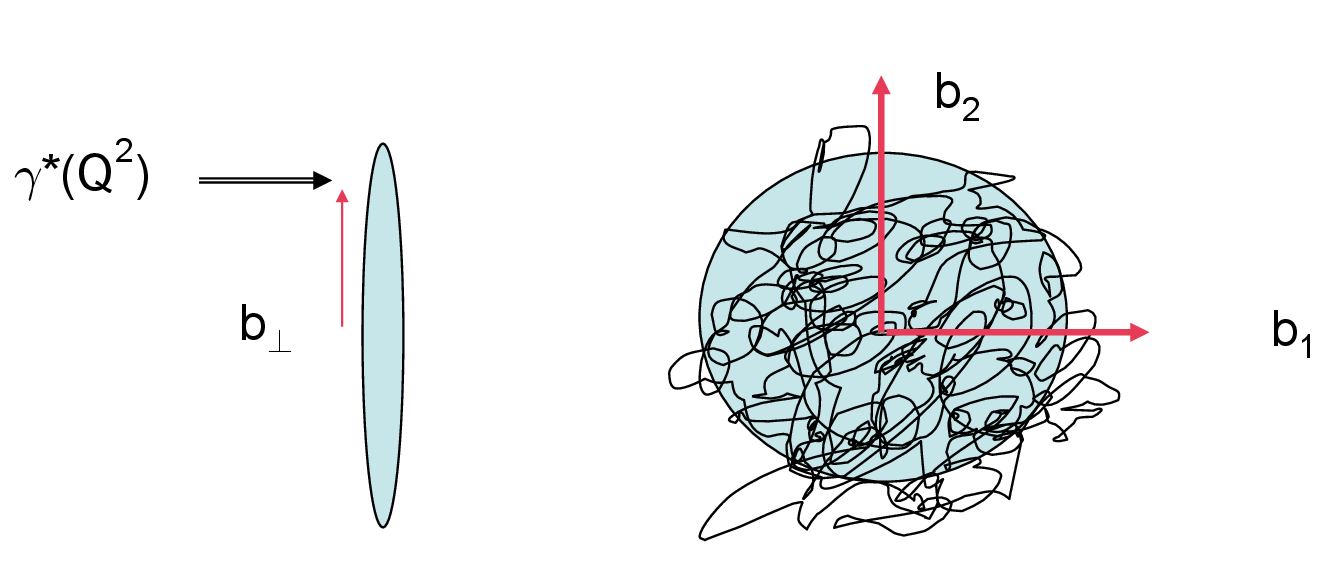}
\qquad
\includegraphics[height = 0.15\textwidth,width = 0.25\textwidth]{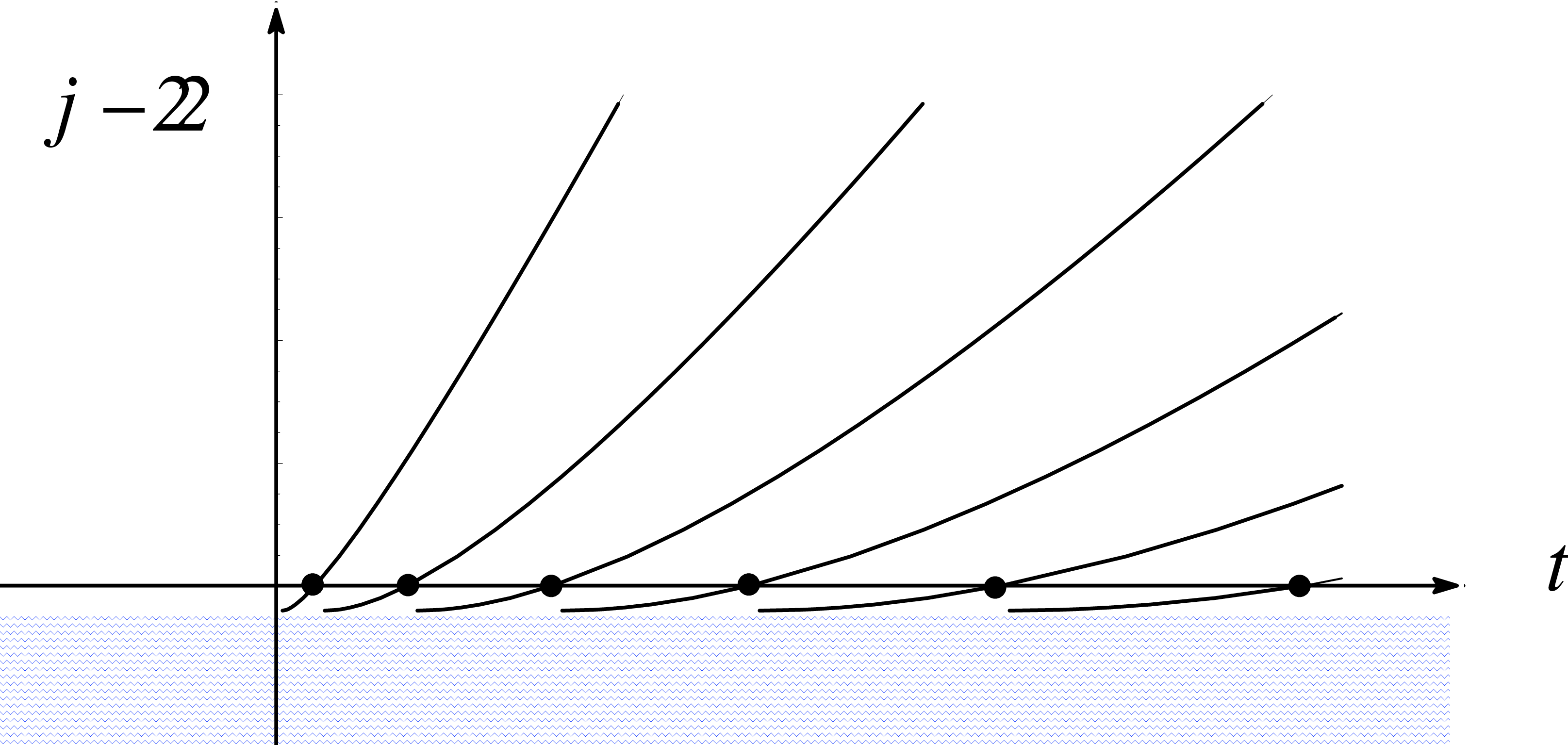}
\qquad
\includegraphics[height = 0.15\textwidth,width = 0.2\textwidth]{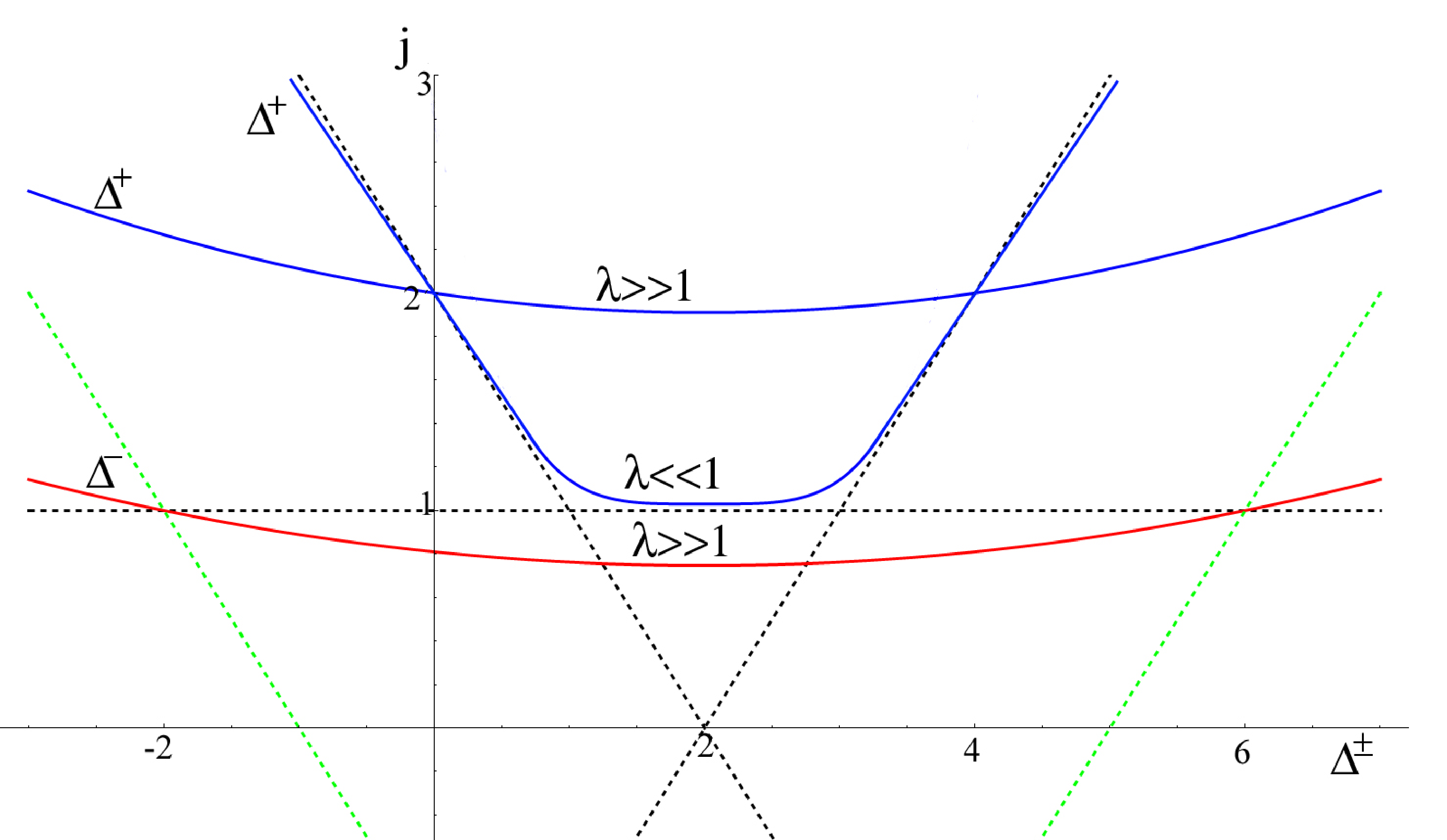}
\quad

 \caption{(a) Intuitive picture for $AdS^5$ kinematics. (b) Schematic representation of $J$-plane singularity structure. (c) Schematic form of $\Delta$-$j$ relation for  $\lambda<<1 $ and $\lambda>>1$ for $C=+1$ and $\lambda>>1$ for $C=-1$.}
        \label{fig:comparison}
\end{figure}

The fact that  5-dim description enters in high energy collision can be understood  as follows. In addition to the usual  LC momenta, $p_{\pm}=p^0\pm p^z$ (2d), and transverse impact variables, $\vec b$ (2d), there is  one more ``dimension": a ``resolution" scale specified by a probe, e.g., $1/Q^2$ of virtual photon in DIS, (see  Fig. \ref{fig:comparison}a.) Because of conformal symmetry, these 5 coordinates transform into each others, leaving the system invariant. In the strong coupling limit, conformal symmetry is  realized as the $SL(2,C)$ isometries of Euclidean $AdS_3$ subspace of $AdS_5$, where $r$ can be identified with $Q^2$.

The traditional description of high-energy small-angle scattering in
QCD has two components --- a soft Pomeron Regge pole associated with  exchanging  tensor
glueballs, and a hard BFKL Pomeron at weak coupling.
On the basis of gauge/string duality,  a coherent treatment
of the Pomeron was provided \cite{Brower:2006ea}.   
These  results agree with expectations for the
BFKL Pomeron at negative $t$, and with the expected glueball spectrum
at positive $t$, but provide a framework in which they are unified \cite{levintan}.

  One important step in  formulating the dual Pomeron involves the demonstration 
\cite{Polchinski:2001tt} that in exclusive hadron scattering, the dual
string theory amplitudes at wide angle, due to the red-shifted local-momenta, $s\rightarrow \tilde s= (R/r)^2 s$ and $t\rightarrow \tilde t= (R/r)^2 t$, give the power laws that are
expected in a gauge theory.   It was also noted that at large $s$
and small $t$ that the classic Regge form of the scattering amplitude should be present in certain kinematic regimes~\cite{Polchinski:2001tt,Brower:2002er}.     Equally important is the fact  that,  with confinement, transverse fluctuations of the metric tensor $G_{MN}$ in $AdS$ acquire a mass and can be identified with a tensor glueball~\cite{Brower:1999nj,Brower:2000rp}. It was suggested in \cite{Brower:2000rp} that, at finite $\lambda$,   this will  lead to a Pomeron with an intercept below 2. That  is, Pomeron can be considered as a  {\em Reggeized Massive Graviton}.

 The {\em dual Pomeron} was
subsequently identified as a well-defined feature of the curved-space string
theory~\cite{Brower:2006ea}. For a conformal theory in the large $N_c$ limit, a dual Pomeron can always be identified with the leading eigenvalue of a Lorentz boost generator
$M_{+-}$ of the conformal group  \cite{Brower:2007xg}.  
The problem reduces to finding the spectrum of
a single $J$-plane Schr\"odinger operator. 
In the strong coupling limit, conformal
symmetry  requires that the leading $C=+1$ Regge
singularity is  a fixed $J$-plane cut.   For ultraviolet-conformal
theories with confinement deformation,  the spectrum exhibits a set of Regge trajectories at
positive $t$, and a leading $J$-plane cut for negative $t$, the
cross-over point being model-dependent.  (See Fig. \ref{fig:comparison}b.) For theories with
logarithmically-running couplings, one instead finds a discrete
spectrum of poles at all $t$, with a set of slowly-varying and closely-spaced
poles at negative $t$.

\section{Conformal Pomeron, Odderon and Analyticity}

At high-energy, analyticity and crossing lead to $C=\pm 1$ vacuum exchanges,  the {\em Pomeron} and  the {\em Odderon}.    The qualitative picture for Pomeron exchange in weak coupling~\cite{Lipatov:1976zz,Kuraev:1977fs,BL} has been understood for a long time, in leading order expansion 
in $g^2_{YM}$ and all order sum in $g^2_{YM} log(s/s_0)$. In the conformal limit,
both the weak-coupling BFKL Pomeron and Odderons correspond to  $J$-plane branch points, e.g.,  the BFKL Pomeoron is a cut at $
j^{(+)}_0$, above $j=1$.    Two leading Odderons have been identified.  Both are branch cuts in the $J$-plane.
One has an intercept slightly below 1
\cite{Janik:1998xj,Braun:1998fs}, and the second has an
intercept precisely at 1  \cite{Bartels:1999yt}. 
(See \cite{BDT,Ewerz:2005rg} for more references. See also \cite{Hatta:2005as,Kovchegov:2003dm,Kovner:2005qj}.)  These are summarized in Table~\ref{tab:intercepts}.  

In the strong coupling limit, conformal
symmetry
dictates that the leading $C=+1$ Regge
singularity is  a fixed $J$-plane cut at   
$
j^{(+)}_0 = 2 - 2/\sqrt{\lambda}+O(1/\lambda).
$
As $\lambda$ increases, the ``conformal Pomeron" moves  to $j=2$ from below, approaching    the  $AdS$  graviton.  We  have recently shown \cite{BDT}   that the strong coupling  {\em conformal odderons} are again
fixed cuts in the $J$-plane, with intercepts specified by the
AdS mass squared, $m^2_{AdS}$, for Kalb-Ramond fields ~\cite{Kalb:1974yc},
\be
j_0^{(-)}=1- m^2_{AdS}/2\sqrt{\lambda} + O(1/\lambda)\;.  \label{eq:dualconformalOdderon}
\ee
 Interestingly, two leading {\em dual odderons} can be identified, parallel the weak-coupling situation. One solution has $m^2_{AdS, (1)} = 16$. There is also  a second solution where $m^2_{AdS, (2)}=0$.   For both cases, they approach $j=1$ in the limit $\lambda \rightarrow \infty$.  We outline below how these features emerge in {\em Gauge/String duality}.

\begin{table}[h]
\begin{center}
\begin{tabular*}{150mm}{@{\extracolsep\fill}||c||l|ll||}
\hline
\hline
 & \multicolumn{1}{c|}{Weak Coupling}   &
 \multicolumn{1}{c}{Strong Coupling}          &    \\
\hline\hline\hline
$\;\;\; C=+1\;\;\;$ & $j^{(+)}_0 = 1
+  ( \ln 2) \; \lambda/ {\pi^2}  +O(\lambda^2)$    
   & $j^{(+)}_0 = 2 - 2/\sqrt{\lambda}+O(1/\lambda)$            &        \\
\hline\hline
$C=-1$    &   $j^{(-)}_{0,(1)} \simeq  1- 0.24717\;  \lambda/\pi + O(\lambda^2)$                                      
     & $j^{(-)}_{0,(1)}=1- 8/\sqrt{\lambda} + O(1/\lambda)$         &               \\
       &     $ j^{(-)}_{0,(2)} = 1 + O(\lambda^3)$              
      & $j^{(-)} _{0,(2)}=1+ O(1/\lambda)$                 &     \\
\hline
\hline
\end{tabular*}
\caption{Pomeron and Odderon intercepts at weak and strong coupling, with  $\lambda = g_{YM}^2 N_c$ the 't Hooft coupling. }\label{tab:intercepts}
\end{center}
\end{table}

\subsection{Flat-Space Expectation for $C=\pm 1$ Sectors}

String scattering  in 10-d flat-space at high energy leads to  a crossing-even  and crossing-odd amplitudes, 
\be
 {\cal T}^{(\pm)}_{10}(s,  t) \to f^{(\pm)} (\alpha'  t)  (\alpha'  s)^{\alpha_\pm(t)} \;,
 \ee
where   $\alpha_{+} (t) = 2 + \alpha'  t /2$ and $ \alpha_{-}(t) = 1 + \alpha'  t /2$ respectively. That is, at $t=0$, a massless state with integral spin is being exchanged, e.g., for $C=+1$, one is  exchanging a massless spin-2 particle, the ubiquitous graviton.  Of course, the coefficient functions, $ f^{(\pm)} (\alpha'  t) $, are process-dependent.

Massless modes of a closed string theory  can be identified with transverse fluctuations coming from a left-moving and a right-moving level-one oscillators, e.g., states created by applying $  a^\dagger_{1,I}\tilde a_{1,J}^\dagger$ to the  vacuum, i.e., $ a^\dagger_{1,I}\tilde a_{1,J}^\dagger |0;k^+,k_\perp\rangle$,   with $k^2=0$.     Since a 10-dim closed string theory in the low-energy limit  becomes 10-dim gravity; these  modes  can be identified  with fluctuations of the metric $G_{MN}$, the anti-symmetric \underline {Kalb-Ramond} background $B_{MN}$ ~\cite{Kalb:1974yc},  and the dilaton, $\phi$, respectively.    It is  important to note that we will soon focus on $AdS^5$, i.e., one is effectively working at $D=5$. With $D=5$,  the independent components for $G_{MN}$ and $B_{MN}$ are 5 and 3 respectively, precisely that necessary for having  (massive) states with  spin 2 and 1 \cite{Brower:2000rp}. For oriented strings, it can be shown that  the symmetric tensor contributes to $C=+1$ and the anti-symmetric tensor contributes to $C=-1$.

\subsection{Diffusion in AdS for Pomeron and Odderon}
\label{sec:diffusion}

 Let us next  introduce diffusion in  AdS.  We will  restrict ourselves  to  the conformal limit.   Regge behavior is intrinsically non-local in the transverse space.  For flat-space scattering in 4-dimension, the transverse space is the 2-dimensional  impact parameter space, $\vec b$. In the Regge limit of $s$ large and  $t<0$, the momentum transfer is transverse. Going to the $\vec b$-space, $ t   \to   \nabla_b^2\;, $
and the flat-space Regge propagator, for both $C=\pm 1 $ sectors,  is nothing but a diffusion kernel,
$  \langle \; \vec b \;|\;(\alpha' s)^{\alpha_{\pm}(0)+\alpha' t\nabla_b^2/2} \;| \; \vec b' \;\rangle $, with  $\alpha_+(0)=2$ and $\alpha_-(0)=1$ respectively. 
In moving to a
ten-dimensional momentum transfer $\tilde t$,   we must keep a term  coming from the momentum transfer in the six transverse directions.  This extra term leads to diffusion in  extra-directions, i.e., for $C=+1$, 
$
\alpha'  \tilde t \to  {\alpha'} \Delta_{P} \equiv
\frac{\alpha' R^2 }{r^2} \nabla_b^2+ {\alpha'} \Delta_{\perp P}.
$
The transverse Laplacian is proportional to $R^{-2}$, so that the added term
is indeed of order ${\alpha'}/{R^2} = 1/\sqrt{\lambda}$. 
To
obtain the $C=+1$ Regge exponents we will have to diagonalize the differential
operator $\Delta_{P}$.  Using a Mellin transform,  $\int_0^{\infty} d\tilde s \; {\tilde s}^{-j-1}$,  the Regge propagator can be expressed as 
$
\tilde s^{ 2+ \alpha' \tilde t /2 }
 = \int \frac{d j}{2\pi i} \; {\tilde s}^j \;    G^{(+)}(j) =  \int \frac{d j}{2\pi i} \; \frac{{\tilde s}^j }  { j-  2 -\alpha' \Delta_P /2 }
$
where  $\Delta_P \simeq  \Delta_j  $,  the tensorial Laplacian.  Using a spectral analysis, it leads to a $J$-plane cut at $j_0^{(+)}$.  

A similar analysis can next be carried out for the $C=-1$ sector. We simply replace the  Regge kernel by 
$
\tilde s^{ 1+ \alpha' \tilde t /2 } =\int \frac{d j}{2\pi i} \; {\tilde s}^j \;    G^{(-)}(j) =   \int \frac{d j}{2\pi i} \; {{\tilde s}^j }   {( j-  1 -\alpha' \Delta_O /2)^{-1} }
$. The operator $\Delta_O(j)$ can be fixed by examining  the EOM at $j=1$ for the associated super-gravity fluctuations responsible for this exchange, i.e., the anti-symmetric Kalb-Ramond fields, $B_{MN}$.  One finds two solutions,
\be
G^{(-)}(j) = \frac{1}{[ j-1 -(\alpha'/2R^2) (\square_{Maxwell} - m^2_{AdS,i}) ]}\; ,
\ee
$i=1,2$, where $\square_{Maxwell}$ stands for the Maxwell operator.  Two allowed values  are
$m^2_{AdS,1}=16 $ and  $m^2_{AdS,2} = 0.$  
 A standard spectral analysis then lead to a branch-cut at $j_0^{(-)}$, given by Eq. (\ref{eq:dualconformalOdderon}).

\subsection{Regge and DGLAP Connection}
 It is also useful to explore the conformal invariance as the  isometry of  transverse $AdS_3$. Upon taking a two-dimensional Fourier transform with respect to $q_\perp$,  where  $t=-q_\perp^2$, one finds  that $G^{(\pm)}$ can be expressed simply as
\be
G^{(\pm)}(z,x^\perp,z',x'^{\perp}; j)  = \frac{1}{4\pi zz' } \frac{ e^{ (2 -\Delta^{(\pm)}(j))\xi}}{  \sinh \xi} \; ,
\label{eq:adschordal}
\ee
where $\cosh \xi = 1+ v$,   $v= [(x^\perp-x'^{\perp})^2+(z-z')^2]/(2 zz')$ 
the $AdS_3$ chordal distance, and $z=R^2/r$,  and 
$
\Delta^{(\pm)} (j) =  2 + \sqrt {2}\;  \lambda^{1/4}
\sqrt{ (j-j^{(\pm)}_0) }
$
is a  $J$-dependent effective  $AdS_5$ conformal dimension \cite{Brower:2006ea,Brower:2007xg,BDT}. The $\Delta-j$ curve  for $\Delta^{(\pm)}$ is shown  in Fig. \ref{fig:comparison}c. A related discussion on $\Delta(j)$  can be found  in \cite{Hofman:2008ar}.

For completeness, we note that,  for both $C=+1$ and $C=-1$, 
it is  useful to  introduce Pomeron and Odderon kernels in a mixed-representation,
\be
{\cal K}^{(\pm)} (s, z,x^\perp, z',x'^\perp) \sim \left(\frac{(zz')^2}{R^4}\right) \int \frac{dj}{2\pi i}   \;     \left[ \frac{(-\tilde s)^j \pm (\tilde s)^j}{\sin \pi j} \right] G^{(\pm)} (z,x^\perp,z',x'^\perp;j)  \;. \label{eq:PomeronOdderonkernel}
\ee
  To obtain scattering amplitudes, we simply fold these kernels with external wave functions. Eq. (\ref{eq:PomeronOdderonkernel}) also serves as the starting point for eikonalization.

\section{Unitarity, Absorption, Saturation and the Eikonal Sum}
\label{sec:eikonal}

 For simplicity, we will focus here on the $C=+1$ sector, assuming all crossing odd amplitudes vanish. It has been shown in 
 Refs.~\cite{Brower:2007qh,Brower:2007xg,Cor} that,
  in the strong coupling limit, a 2-to-2 amplitude, $A(s,t)$, in the near-forward limit  can be expressed in
  terms of a ``generalized'' eikonal representation,
\be
\; A_{2\to2}(s,t) =\int dz dz' P_{13}(z) P_{24}(z') \int d^2b \; e^{-ib^\perp q_\perp}  \widetilde A(s,b^\perp,z,z')\;,
\ee
where 
$
\widetilde A(s,b^\perp,z,z')= 2 i s \left [1- e^{i\chi(s,b^\perp, z,z')} \right] \;,
$
and  $b^\perp = x^\perp - x'^\perp$ due to
translational invariance.    The probability distributions for left-moving, $P_{13}(z)$, and right moving, $P_{14}(z)$ particles are products of initial (in) and final (out)
particle wave functions. The eikonal, $ \chi$,  can be related to  the strong coupling Pomeron kernel \cite{Brower:2006ea,Brower:2007xg}, and can be expressed  as  the inverse Mellin transform of $G^{(+)}(j,x^\perp - x'^\perp,z,z')$.

We note the  salient feature of eikonal scattering locally in transverse
$AdS_3$, and the near-forward field-theoretic amplitude is obtained
from a bulk eikonal amplitude after convolution. It is useful to focus
our attention on the properties of the bulk eikonal formula
$\widetilde A(s,b^\perp,z,z')$ itself. For $\chi$
real, it is elastic unitary.  On the other hand, when $\chi$ is
complex, (with ${\rm Im} \chi >0$), one has inelastic production. Absorption and saturation can now be addressed in this context. It is also important to note that, for Froissart bound, confinement is crucial.   Discussion on these and related  issues  can be found in Ref. \cite{Brower:2007xg}. 

We end by pointing out one unique feature of strong coupling -- the eikonal is predominantly real. To simplify the discussion, let us consider the second order contributions to the imaginary part of the elastic amplitude.  The AGK cutting rule for the imaginary part of the elastic amplitude  generalizes to
\be
\cos (j_0\pi) |\chi|^2 = \left[ 1 -2 \sin^2(j_0 \pi/2) -2 \sin^2(j_0 \pi/2)  + 2 \sin^2(j_0 \pi/2) \right] |\chi|^2
\ee
where the first term on the right is due to the elastic scattering, the last term is due to two-cut-Pomeron contribution, and the second and the third are due to one-cut-Pomeron  contributions. 
The tradition weak coupling approach to diffraction scattering has $j_0\simeq 1$, leading to a net negative contribution: $-1 =  1 -2 -2 +2$.
This leads to absorption, already dominant at second order. However, for extreme strong coupling, one has $j_0\simeq 2$, leading to a positive cut contribution: $1 =  1-0-0 + 0$.  This is consistent with  scattering being predominantly elastic. However, the real world is neither strictly weak coupling nor strong coupling. For $j_0\simeq 1.5$, one finds the two-Pomeron contribution vanishes: $ 0 =  1 -1 -1 +1$.
That is, what used to be the dominant correction to elastic scattering now vanishes. Clearly, these issues deserve further examination.  For recent applications of \cite{Brower:2006ea,Brower:2007qh,Brower:2007xg,BDT,Cornalba:2006xk,Cornalba:2006xm,Cor}  for DIS, see \cite{Levin:2008vj,HIM,Cornalba:2008sp}.

%------------------------------------------------------------------------------
%       Bibliography
%------------------------------------------------------------------------------

\bibliographystyle{utphys}
\bibliography{ismd08}

\end{document}